# Directional Measurements in Urban Street Canyons from Macro Rooftop Sites at 28 GHz for 90% Outdoor Coverage

Jinfeng Du, *Member, IEEE,* Dmitry Chizhik, *Fellow, IEEE,* Reinaldo A. Valenzuela, *Fellow, IEEE,*
Rodolfo Feick, *Life Senior Member, IEEE,* Guillermo Castro, Mauricio Rodriguez, *Senior Member, IEEE,* Tingjun Chen, Manav Kohli, and Gil Zussman, *Senior Member, IEEE*

*Abstract* Path gain and effective directional gain in azimuth in urban canyons from actual rooftop base station sites are characterized based on a massive data set of 3000 links on 12 streets in two cities, with over 21 million individual continuous wave power measurements at 28 GHz using vertically polarized antennas. Large street-to-street path gain variation is found, with median street path gain varying over 30 dB at similar distances. Coverage in the street directly illuminated by a roof edge antenna is found to suffer an average excess loss of 11 dB relative to free space at 200 m, with empirical slope-intercept fit model representing the data with 7.1 dB standard deviation. Offsetting the base antenna 5 m away from roof edge, as is common in macro cellular deployments, introduces an additional average loss of 15 dB at 100 m, but this additional loss reduces with distance. Around the corner loss is well modeled by a diffraction formula with an empirically obtained diffraction coefficient. Effective azimuthal gain degradation due to scatter is limited to 2 dB for 90% of data, supporting effective use of high gain antennas in urban street canyons.

## I. INTRODUCTION

Large amounts of spectrum available at mm wave bands promise large communication capacities, provided adequate link budget can be maintained despite increased propagation losses. Link budget can generally be improved both through deploying a dense cell network and using directional antennas for higher directional gain. Since the cost of deploying a network increases with cell site density, it is critical to determine coverage range from each site to allow assessment of commercial viability in environments of interest. Since effective directional antenna gain is reduced by scattering, it is important to quantify the achievable directional gain in realistic environments.

Numerous measurement campaigns [1]-[19] and channel modeling using ray-tracing simulations have been carried out in the past few years aiming for fundamental understanding of mm wave propagation in dense urban environments featuring street canyons.

Uncertainty in predicting average local path gain at a location is due to both location-location variability (a.k.a. shadow fading) as well as uncertainty due to estimation of model parameters, i.e. slope and intercept, based on data fit. High reliability outage statistics (e.g., 90% coverage), require hundreds of links for a particular environment to make sure the model uncertainty is much less than the RMS spread in the data. That requires many hundreds of link measurements, as opposed to dozens. Very extensive measurements of path gain at 3.35, 8.5, and 15.75 GHz were reported for a single Tokyo street in [5] using a base station at 4 m and a terminal at 1.6 and 2.7 m. A two-ray model with an empirically determined ground height was found to be effective at modeling path loss for the 2.7 m terminal while a single-slope model did well for 1.6 m. Similarly, low-height measurements in [26] found two-ray was effective for lamppost heights, while a blockage model was needed for peer-peer links. While such measurements are very useful in characterizing peer-peer and lamppost coverage, there is strong commercial motivation in characterizing coverage from rooftop macro cellular sites. Two high

---
Jinfeng Du, Dmitry Chizhik and Reinaldo. A. Valenzuela are with Nokia Bell Labs, Crawford Hill, Holmdel, NJ 07733, USA (e-mail: jinfeng.du@nokia-bell-labs.com, dmitry.chizhik@nokia-bell-labs.com, reinaldo.valenzuela@nokia-bell-labs.com).

Rodolfo Feick is with Department of Electronics and Centro Científico y Tecnológico Valparaíso-CCTVal, Universidad Técnica Federico Santa María, Valparaíso, Chile (e-mail: rodolfo.feick@usm.cl).

Guillermo Castro and Mauricio Rodriguez are with Escuela de Ingeniería Eléctrica de la Pontificia Universidad Católica de Valparaíso, Valparaíso, Chile (email: mauricio.rodriguez.g@pucv.cl, guillermo.castro@pucv.cl).

Tingjun Chen is with the Department of Electrical Engineering, Duke University, Durham, NC 27708, USA. He was with the Department of Electrical Engineering, Columbia University, New York, NY 10027, USA (email: tingjun.chen@duke.edu).

Manav Kohli and Gil Zussman are with Department of Electrical Engineering, Columbia University, New York, NY 10027, USA (email: mpk2138@columbia.edu, gil.zussman@columbia.edu).

base station locations were used in [2] to collect wideband directional data for 43 terminal locations. Ultra-wideband measurements [30][31] at 30 and 60 GHz were collected from two street canyons using dual-polarized horn antennas scanning at steps of 30°, where the transmitter was placed about 17 m high and the receiver moved to a dozen locations along the street for a distance up to 140 m, where both canyons ended with a building. Such measurements are useful for an initial survey, but a reliable quantitative model requires hundreds of locations, as mentioned above.

Around-corner propagation along street canyons is of special importance for coverage planning and inter-cell interference management. Various path loss models have been developed in the literature to characterize around-corner propagation along urban street canyons. Traditional urban macro (UMa) NLOS models with exponent close to 4, such as [20][21], are designed for scenarios where over-the-top propagation is dominant. For cases where both terminals are below average building height, such as in Manhattan street canyon with wall or lamppost mounted base stations, around-corner propagation is dominant for NLOS coverage. For sub-6 GHz frequencies (e.g. 430 MHz to 4860 MHz in [25]), street canyon NLOS models with Manhattan grid cell layout [21][22][25] have been extended to mm wave bands in [9] using ray tracing simulation over regular street grid, and in [10][17] using field measurements. In [10] the ITU-R model [22] was fitted using urban street canyon measurements from multiple cities in multiple frequency bands, and the NLOS distance exponent was found to be around 3 over multiple bands. In [17] the ITU-R model [25] was fitted using 28 GHz around-corner NLOS measurements with two different distances from the corner, and the path loss exponents were found to be around 4 in one case and around 11 in the other. A dual-slope model, which is essentially a simplification of ITU-R model [22][25] has been adopted in [9] using the LOS path to the corner as reference distance and the unwrapped route distance in path loss calculation for NLOS segment.

The principal objective here is to characterize reliably (with empirical model uncertainty much smaller than shadow fading uncertainty) mm wave coverage from macro base stations to same-street outdoor terminals both for roof-edge and offset from edge base antennas. This is done to assess improvement in rate offered by mm wave spectrum over traditional microwave band cellular coverage in more limited bandwidth. Our work derives its conclusions based on over 3000 continuous wave (CW at 28 GHz) links measured at multiple base locations, collected from 12 streets in Manhattan, NY and Valparaíso, Chile. There are in total over 21 million individual CW power samples. Each link measurement consisted of over 30 azimuth scans. Slope-intercept fit represents measured path loss in the street canyons with a RMS deviation of 7.1 dB. Median degradation suffered by offsetting the base antenna from the roof edge, as is often done in practical installations, is found to be 15 dB at 100 m. We find high effective directional gains are available even in the presence of street-induced scatter, with 90% of locations suffering under 2 dB gain reduction. Path gain for around-corner propagation is well characterized using a diffraction-based model with empirically-obtained coefficients. The large data set allows statement of empirical models with 90% confidence interval of under 1 dB for path gain and under 0.5 dB for effective directional gain distribution. The resulting models are used to predict achievable rates, which are found to exceed 300 Mbps for 90% of outdoor locations for 12 sites/sq km with 400 m inter-site-distance and 800 MHz bandwidth at 28 GHz.

## II. MEASUREMENT DESCRIPTION

### A. Measurement equipment

To maximize link budget and data collection speed, we constructed a narrowband sounder, transmitting a 28 GHz CW tone at 22 dBm into an omnidirectional antenna (2 dBi, in Manhattan) or a 55° horn (10 dBi, in Valparaiso). The signal received by the 10° (24 dBi) horn, rotating in azimuth to collect signals arriving from all directions, was amplified by several adjustable gain low-noise amplifiers (5 dB effective noise figure), mixed with a local oscillator, resulting in an IF signal centered at 100 MHz with effective bandwidth of 20 kHz, whose power was measured and converted to digital values with a power meter and stored on a computer. Both horn antennas were vertically polarized, with cross-pol isolation over 25 dB. The complete receiver, including the data acquisition computer, was mounted on a rotating platform allowing a full angular scan at speeds up to 300 rpm. Given the sampling rate of 740 power samples/sec, at 300 rpm, a power measurement was captured every 2.5 degrees. This angular sampling is substantially finer than the 10° beamwidth of the spinning receive horn.

The system was calibrated in the lab and anechoic chamber to assure absolute power accuracy of 0.15 dB. The full dynamic range of the receiver (from noise floor to 1 dB compression point) was found to be 50 dB, extensible to 75 dB using switchable receiver amplifiers. In combination with removable transmit attenuators (0 to 40 dB, used at very short ranges), measurable path loss allowing at least 10 dB SNR ranged from 61 dB (1 meter in free space) to 137 dB (e.g. 200 m range with 30 dB excess loss). Measurable path loss extends to 171 dB with directional antenna gains. This follows from 32 dBm EIRP, 24 dBi receive horn gain, 75 dB maximum effective receiver gain (combined LNA/ mixer) and target receive power of -40 dBm to be 10 dB above the -50 dBm noise floor.

*B. Measurement environment*

To emulate canyon coverage from a rooftop base, the spinning horn receiver was placed at roof edges at multiple commercial base station (BS) heights (as shown in Fig. 1a). Measurements were done while placing the rooftop base in view of the street canyon as well as base antenna offset away from the edge of the roof, as in many commercial sites. The transmitter, either an omnidirectional antenna (in Manhattan) or a 55° horn (in Valparaíso), was placed on a tripod in the middle of a sidewalk (Fig. 1b). Transmitter placements included both the street in view of the base receiver (same-street) as well as around-the-corner cases. Same-street coverage from lamppost-mounted base was also assessed.

Over 3000 links were measured for dense urban macro deployments, corresponding to over 21 million individual power measurements. Measurements were done from 8 buildings, covering 12 streets, 8 in Manhattan and 4 in Valparaíso, with BS height varying from 15 to 51 m and transmitter-receiver separation from 35 to 800 m, measured every 3 to 6 m.

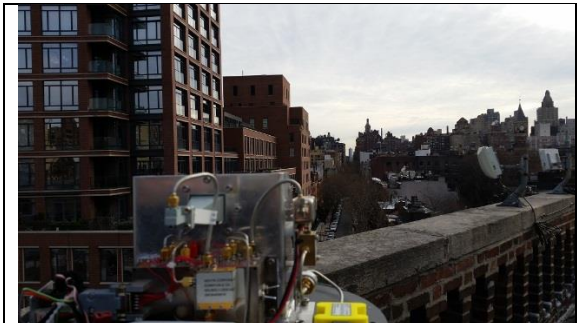

Fig. 1a. Spinning horn receiver overlooking West 11th Street in Manhattan from a roof edge 18 m high above ground.

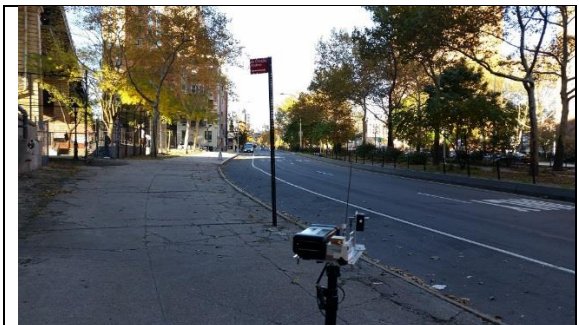

Fig. 1b. Omnidirectional transmitter placed in the middle of a sidewalk.

*C. Data processing*

The azimuthal average of received power over all angular directions, denoted as $P_{all}$, has been shown to be equivalent to the average omnidirectional power (see detailed derivation given by equation (1)--(6) in [23]). We can therefore estimate the effective omnidirectional path gain $P_G$ from $P_{all}$ by subtracting transmit power $P_T$, nominal transmit antenna gain $G_T$ and nominal elevation gain of the receive antenna $G_{elev}$:

$$P_G = 10\log_{10}(P_{all}) - P_T - G_T - G_{elev} \quad [\text{dB}]$$

Both transmit antenna gain and receive elevation gain are assumed undegraded by scattering. Such idealization is justified by the observation that the transmit antenna azimuth beamwidth (55° for the horn and 360° for omni) employed in measurements is wider than the expected angle spread. Similar justification is made for the elevation gain, since the elevation angular spread has been reported as being small in channel models such as 3GPP 38.901 [20].

## III. PATH GAIN

Same-street measurements were collected with an outdoor terminal placed at successive positions along the street (every 3 to 6 m) in the middle of a sidewalk, and a (spinning horn) receive antenna either at roof edge directly illuminating that street or, else, offset by about 5 m from roof edge. Illustration of the two scenarios is in Fig. 2.

As the distance between the transmitter and the receiver increases, the first Fresnel zone radius increases from a few tens of centimeters to over a meter at 500 m and beyond. We may have a visual LOS path between the transmitter and the receiver but the presence of clutter in the first Fresnel zone can cause substantial loss. To avoid confusion, we use "same-street" to label links that would have been in LOS condition if there were no street clutter, and use "around-corner" for NLOS links collected from perpendicular streets where the direct paths are blocked by buildings.

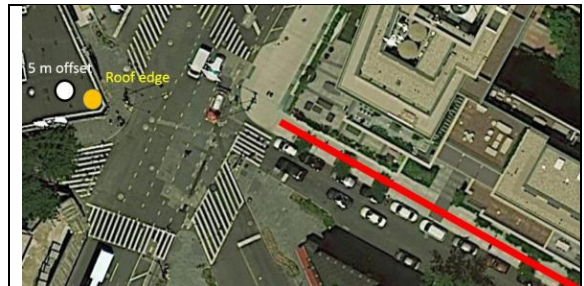

Figure 2. Illustration of roof-edge (orange) and offset from roof edge (white), base antenna placement (on left). Same-street terminal is moved along the red line on right.

## A. Same-street coverage from roof edge antenna

A typical measurement consisted of placing the rotating 10° receive horn (BS) at the edge of a roof of a building and placing a transmitter (omnidirectional or a 10 dBi horn manually re-aimed towards the receiver from each transmitter location) on a tripod 1.5 m high at different ranges along a sidewalk, mimicking a user equipment (UE). No effort was done to include or exclude blockage by street clutter, such as vegetation, vehicles, pedestrians, scaffolding. The intent is to emulate coverage of street in the presence of such obstructions. In total 1650 same-street links were measured on 12 streets (Manhattan and Valparaíso), and each link recorded power samples for at least 10 seconds (over 7400 samples). All such path gain vs. distance results are shown in Fig. 3.

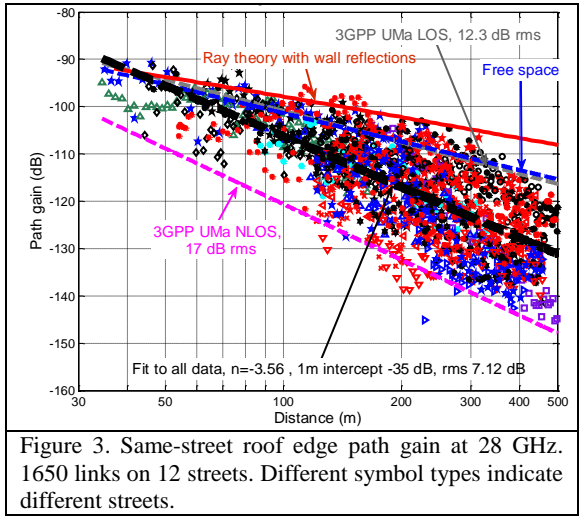

Figure 3. Same-street roof edge path gain at 28 GHz. 1650 links on 12 streets. Different symbol types indicate different streets.

Slope-intercept fit to the measured path gain with respect to distance, including 90% confidence intervals for both parameters for all the roof-edge data is

$$P_{Edge} = A + 10n\log_{10} d + N(0,\sigma),\ \sigma = 7.1\ \text{dB} \\ A = -35.0\ +/-\ 2.7\ \text{dB},\ n = -3.56 +/-\ 0.12 \quad (1)$$

In (1) $A$ [dB] is the 1-m intercept, $n$ is the distance exponent, $d$ [m] is the distance, and $\sigma$ [dB] is the RMS error. The deviation of measured path gain vs. distance from the linear fit (1) is found to be distributed within 0.4 dB of a log-normal distribution with the same standard deviation, for 99% of points.

At 40 m the fit line is close to Friis free space predictions, but at 200 m there is an excess loss of 11 dB, increasing to 20 dB at 500 m. A fit using a fixed intercept at 1 m (set to be same as Friis) and adjusting distance exponent only, results in distance exponent $n= -2.48$, with rms deviation of 7.5 dB.

Fitting the path gain measurements in 8 street sets in Manhattan and 4 in Valparaíso separately results in fit lines about 2 dB apart. This is much smaller than path gain differences at a fixed range between streets within each city. It is found that the 3GPP UMa LOS model, which is very close to the free space prediction, has a RMS deviation from our data of about 12.3 dB, while the 3GPP UMa NLOS model has 17 dB RMS deviation from data as shown in Fig. 3. As can be seen, these models are, correspondingly stronger/weaker than data collected here by about 12 dB at 200 m. That may be due to environment here being different than environments used to define 3GPP models. All streets measured here contained clutter (vegetation, vehicles, etc.) which may be responsible for the excess loss observed. Yet such losses are not as substantial as "true" NLOS cases, involving blockage by buildings, which is what the 3GPP NLOS model is intended for. The comparison to data fit and 38.901 UMa models is summarized in Table 1.

Table 1. Same-street path gain from roof-edge antenna. Empirical data compared to its own fit and 3GPP models.

| Model | 1-m intercept $A$ [dB] | Distance exponent $n$ | RMS error [dB] |
|---|---|---|---|
| Fit to data | -35.0 | -3.56 | 7.1 |
| 3GPP 38.901 UMa, LOS | -56.9 | -2.20 | 12.3 |
| 3GPP 38.901 UMa, NLOS | -42.5 | -3.91 | 17.0 |

Also shown in Fig. 3 in red is a ray theory prediction for a canyon, which includes up to 10 reflections from the walls as well as from the ground. The canyon walls were represented as vertical planes, with a relative dielectric constant of 5, appropriate for concrete. The ray powers are summed incoherently to produce path gain prediction that is monotonically decreasing with distance. The ray theory prediction of average power is higher than in free space, due to wall and ground reflections. Ray theory is also seen to predict some 13 dB higher power than fit to observations at 200 m. This is attributed to unmodelled street clutter, such as scatter by (generally sparse) trees, lampposts, vehicles and pedestrians. Such objects are generally difficult to represent in ray theory, both in terms of availability of environmental details as well as inapplicability of simple specular reflection models to scatter from complex, rough objects. This is in contrast to reported ray tracing accuracy in [19]. We also note that in [9], fit to ray tracing calculation in LOS without street clutter results in higher power than free space prediction, similar to red curve in Fig. 3, again in contrast to our measurements. We conclude, based on this study, that simple ray tracing is inadequate to represent the data collected here even in the case of a nominally LOS street canyon.

Distributions of measured path gain for individual streets are shown in Fig. 4, each measured at regular intervals at ranges from 30 to 500 m. They are left unlabeled for clarity. At any fixed range the street-street variation is strong, with median gains spanning -102 to -133 dB. This is possibly a consequence of differing amounts of vegetation and building heights.

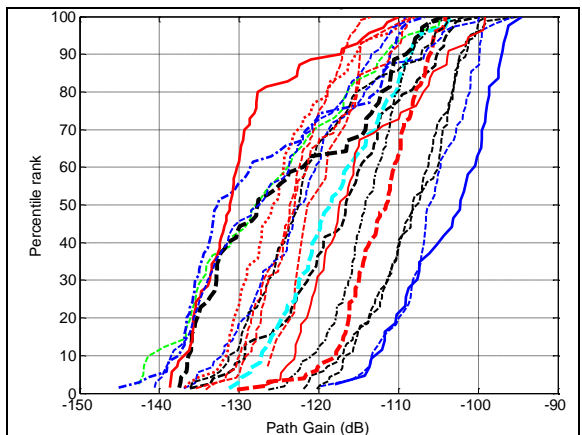

Figure 4. Distributions of measured path gain between a roof-edge Rx and a same-street terminal Tx on 12 streets, showing street-to-street variability.

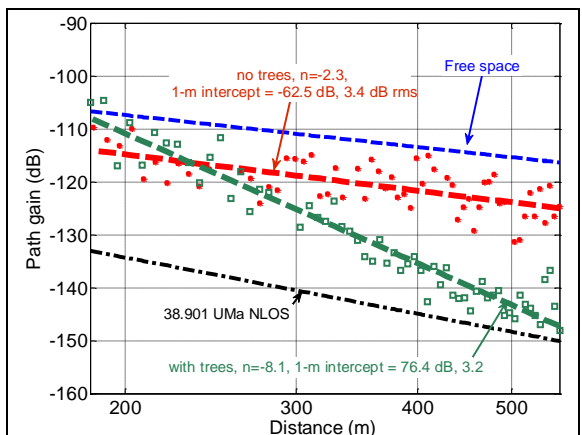

Figure 5. Comparison of path gain on streets with and without trees, measured at the same roof-top site.

Impact of different street environment on measured path gain is illustrated in Fig. 5, containing measurements along 7th Ave in Manhattan (which has practically no trees) and along W 11th Street (with a lot of trees) measured from the same roof-top at the corner of these two streets. The data from both streets has been included in the overall data set shown in Fig. 4. It may be observed that the path gain on the street with trees has a far higher distance exponent of -8.1 as opposed to -2.3 on a street with no trees. This leads to some 23 dB more loss at 500 m. Notably this is so despite the absence of leaves during these winter measurements.

### B. Same-street coverage from rooftop antenna offset from roof edge

Base station antennas are often deployed away from roof edge, closer to the middle of the building, to conceal them from street view based on aesthetic considerations. Naturally, there is concern that street coverage is degraded due to roof blockage, particularly at higher frequencies. To emulate conditions experienced by base station antennas when offset from the roof edge, the spinning horn receiver was placed about 5 meters away from roof edge, as is common, at a height of 1 m above the roof. This caused line of sight to be blocked by a parapet to the terminal locations within 100 m or so, depending on building height, which varied from 19 to 51 m in these offset measurements. Measured path gain for all 1277 such links on 9 streets are shown in Fig. 6, together with a slope-intercept fit (including 90% confidence intervals):

$$P_{\text{offset}} = A + 10n \log_{10} d + N(0,\sigma), \ \sigma = 7.0 \text{ dB} \quad (2)$$
$$A = -94 \ +/- \ 3.7 \text{ dB}, \ n = -1.44 +/- \ 0.16$$

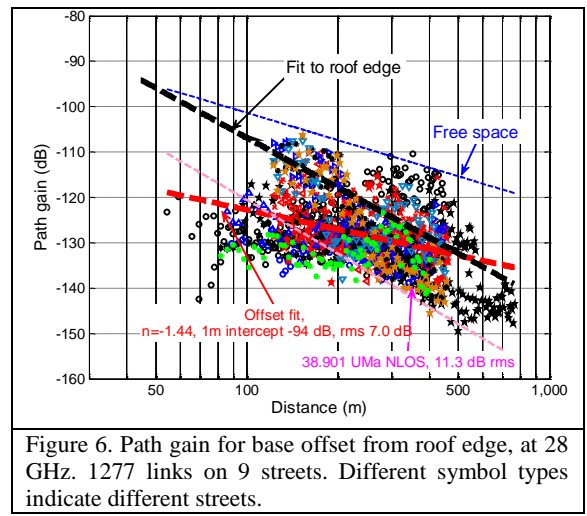

Figure 6. Path gain for base offset from roof edge, at 28 GHz. 1277 links on 9 streets. Different symbol types indicate different streets.

Fitting the measured path gain by adjusting the slope-only, with a fixed intercept set for 1m to the Friis value, results in an exponent of $n= -2.80$, and rms deviation of 7.7 dB.

At ranges of less than 100 m, where the offset base station is blocked from direct view of the street, 25 dB to 50 dB excess loss relative to free space is observed. at ranges beyond 100 m, path loss is about 20 dB worse than free space.

Fit to roof-edge data (from Fig. 3) is also plotted in Fig. 6 for reference. For ranges under 100 m, offsetting the antenna about 5 meters away from roof edge introduces over 15 dB average extra loss. At longer ranges, the difference is reduced. Since the presumed degradation mechanism is diffraction loss at the roof edge, the excess loss becomes smaller at longer ranges where the diffraction angle is small. At very long ranges, the terminal is no longer blocked by the roof edge, despite being offset. In addition, reflections from nearby buildings provide a possibility of non-diffracted paths.

### C. Same-street coverage from lampposts

To assess same-street coverage from lamppost-mounted BS, measurements were collected at 422 links on 3 streets with BS receiver mounted at heights ranging from 8 to 15 m and terminal Tx on the street. The resulting path gain values, shown in Fig. 7, are represented by a slope-intercept fit as:

$$P_{\text{lamppost}} = A + 10n \log_{10} d + N(0, \sigma),$$
$$A = -60.4 \text{ dB}, \; n = -2.42, \; \sigma = 5.5 \text{ dB}$$

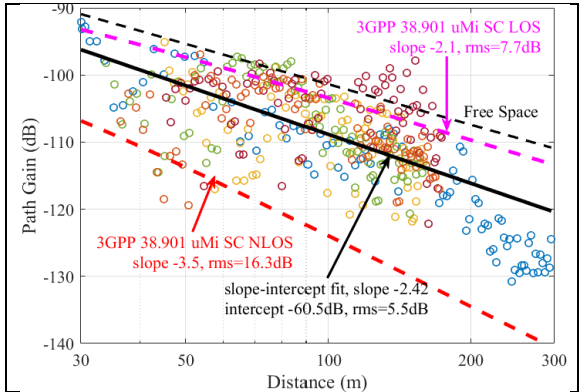

Figure 7. Same-street path gain measured for 422 links with BS on lamppost (8 to 15 m high), terminal on street.

The fitted model shows the path gain in this environment suffers about 9 dB more loss than free space at 200 m. Fitting the lamppost data with an adjustable distance exponent, while fixing the 1-m intercept to its free space value, results in an exponent of $n=-2.37$, and RMS deviation of 5.5 dB. Using the rooftop-derived slope-intercept values (1) to represent lamppost data results in only an increase in RMS deviation to 6.0 dB, suggesting very similar path gain behavior for roof-edge and lamppost mounted base stations in same-street coverage. These lamppost results have been included in the overall street-by-street path gain distributions reported in Fig. 4.

### D. Around-the-corner coverage from roof edge

As illustrated in Fig. 8, when both roof-edge Rx and street-level transmitter are on the same street, they are in "nominally LOS" conditions with possible blockage from trees or street fixtures. As one terminal moves around a corner into a perpendicular street, the propagation channel changes from same-street to being around-one-corner, and then possibly NLOS around-two-corners [9][25]. In those geometry-based models, it is the "Manhattan distance", i.e., the unwrapped distance along the route, that is used in path loss modeling. This is in contrast to traditional models where Euclidean distances are used, which may result in large RMS fitting error (11 dB as reported in [18]) for street canyon NLOS channels.

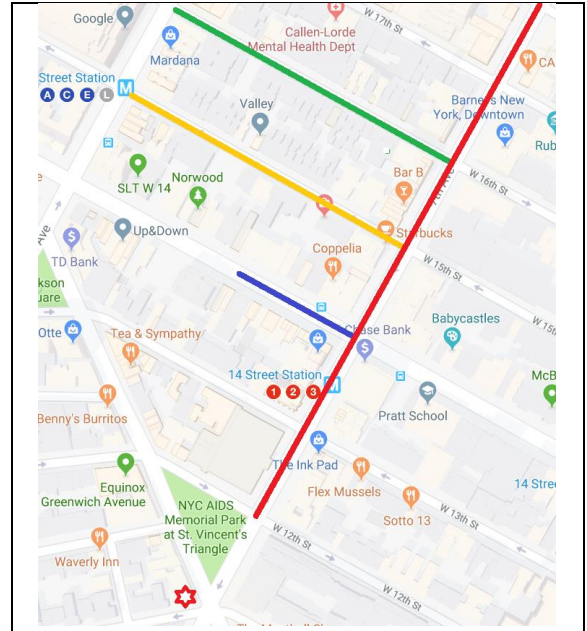

Figure 8. Around-corner measurements with rotating Rx (red hex star near bottom) on the roof of a six-story building, and the 1.5m-high Tx, moving along the same street (red line) and around-corner (blue, orange, and green lines).

Based on the success of modeling around-corner propagation along hallways inside buildings [23], we propose two theory-inspired empirical models, namely, a single-slope scattering model [23] where the corner is treated as a new source, and a single-slope diffraction model where edge-diffraction around the corner is assumed to be the dominant mechanism. Let $x$ be the unwrapped distance between the two terminals, and without loss of generality assuming one terminal is fixed, and the other terminal moves away along the street and then turns into a perpendicular street. As distance $x$ increases, the propagation channel changes from same-street to around-corner.

The diffraction model is given by

$$P_d(x) = \begin{cases} P_1 + 10n \log_{10}(x), & 1 < x < d_c \\ P_1 - \Delta + 5n \log_{10}(d_c(x - d_c)x), & x > d_c \end{cases} \quad (3)$$

where $d_c$ is the distance from the fixed terminal to the corner, $P_1$ is the intercept at 1 m distance, $n$ is the common distance exponent before and after the corner, and $\Delta > 0$ is the empirical "corner loss", replacing diffraction coefficient used in canonical edge diffraction models. When fixing $n=-2$ it is similar to the corner model in [29].

The scattering model is given by

$$P_s(x) = \begin{cases} P_1 + 10n\log_{10}(x), & 1 < x < d_c \\ P_1 - \Delta + 10n\log_{10}(d_c(x-d_c)), & x > d_c \end{cases} \quad (4)$$

It is also worthwhile to compare (3) and (4) with the dual-slope corner model proposed in [9]

$$P(x) = P_1 + 10n_1\log_{10}(x), \quad 1 < x < d_c$$
$$= P_1 + 10n_1\log_{10}(d_c) - \Delta + 10n_2\log_{10}\left(\frac{x}{d_c}\right), \quad x > d_c \quad (5)$$

where $n_1$ and $n_2$ are the distance exponents of before- and after-the-corner segments, respectively.

We validate those models using measurements collected in Manhattan. We placed the rotating horn receiver on the roof of a six-story building located at a street intersection and moved the omnidirectional transmitter (mounted on a 1.5 m high tripod) along the sidewalk of a 30 m wide street. The measurement routes are illustrated in Fig. 8 where the same-street route is indicated by a red line and three around-corner routes in blue, orange, and green, respectively. The surrounding buildings are of various heights, and the ones blocking the direct path are from 10 to over 20 story high, much higher than the six-story roof where the base is placed. A total of 98 same-street links (i.e. before corner) were collected from 91 to 565 m along the street. Around-corner links were collected on three perpendicular streets where the corners are 244 m, 332 m, and 414 m away from the base, and the length of the NLOS paths extend up to 210 m. The measured data and the single-slope diffraction model are shown in Fig. 9 and the fitted parameters of the models are summarized in Table 2. The measured data and the fitted models indicate that after turning around a corner in a Manhattan street canyon, signal drops about 14 dB after 10 m into the corner, and about 21 dB after 50 m into the corner.

The single-slope diffraction inspired model provides the best fit, with 3.4 dB RMS error using only two parameters (slope and corner loss). This suggests that signal re-spreading from the corner are an important propagation mechanism in urban Manhattan street canyons (continuous tall buildings on both sides of streets with no or thin foliage). The dual-slope model (5) provides larger fitting error despite the fact that it uses more parameters (two slopes, corner loss) than the diffraction model (3). Therefore, a diffraction formulation is apparently better for around-the-corner path loss in urban street canyons. The scattering model (4) has the highest RMS error of 6.6 dB. Allowing floating intercept will only slightly reduce the RMS error for diffraction model (3) and dual-slope model (5), but would substantially reduce the RMS error to 4.1 dB for the scattering model (4).

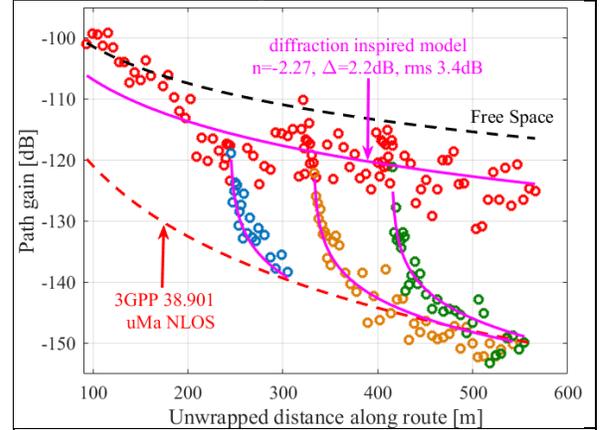

Figure 9. Around-corner measurements with the single-slope diffraction inspired channel model using fixed intercept of Friis @ 1m, with rms fitting error of 3.4 dB.

Table 2. Fitted parameters of candidate channel models for around-corner links, with 1-m intercept either fixed at Friis value or fit to data.

| Model | Intercept $P_1$ | Exponent $n$ | Corner loss $\Delta$ | RMSE [dB] |
|---|---|---|---|---|
| Diffraction model (3) | Friis @1m -61.4 dB | -2.27 | 2.2 dB | 3.4 |
| Scattering model (4) | Friis @1m -61.4 dB | -2.23 | 0 dB | 6.6 |
| Dual-slope model (5) | Friis @1m -61.4 dB | $n_1$= -2.27 $n_2$=-12.3 | 12.0 dB | 4.0 |
| Diffraction model (3) | -52.1 dB | -2.63 | 0 dB | 3.2 |
| Scattering model (4) | -81.3 dB | -1.44 | 0 dB | 4.1 |
| Dual-slope model (5) | -11.8 dB | $n_1$= -3.35 $n_2$=-12.2 | 11.8 dB | 3.6 |

Note that for both diffraction model and the scattering model, the corner loss is very small, only 2.2 dB for diffraction model and 0 dB for scattering model. This is in sharp contrast to indoor corridor around-corner propagation reported in [23]. This may also be compared to the theoretical edge diffraction coefficient, which at large diffraction angles (deep shadow) is on the order of -42 dB at 28 GHz [28]. Similar findings have been reported in around-street corner measurements for 400 MHz to 4.8 GHz [29] where an empirical "scattering coefficient" $S^2$ equivalent to $\Delta$ was obtained through fit to measurements and found to be much larger than the theoretical edge diffraction coefficient. This was attributed to scattering from lampposts and other street furniture. Thus, this level of environmental detail may render ray tracing impractical for mm wave bands.

## IV. EFFECTIVE DIRECTIONAL GAIN IN AZIMUTH

High antenna gain is essential to compensate for the high propagation losses in mm/cm wave bands. However, the potential directional gain degradation caused by angular spread would make it less effective. In this section we quantify the azimuth directional gain degradation experienced by the $10°$ horn at the base[2]. The effective pattern of an antenna, as seen from field power-angular measurement, is the convolution of its nominal pattern (as measured in an anechoic chamber) and the channel angular response (scattering pattern). Channel angular spread widens the effective antenna pattern and therefore reduces its effective gain. In all cases, the azimuthal gain is defined as the ratio of the maximum power to average power over all angles:

$$\text{Azimuth gain} = \frac{\max_\varphi P(\varphi)}{\frac{1}{2\pi}\int_0^{2\pi} d\varphi\, P(\varphi)} \quad (6)$$

It was found that in many streets the effective gain for the roof-edge base generally increased weakly with distance. The implicit angle spread thus decreases with distance in street canyons. An example of this is illustrated by plotting normalized azimuth patterns in Fig. 10, where the blue pattern, measured at 580 m range, shows an unambiguous main lobe, some 40 dB above sidelobes, while the black pattern, measured at 100 m, shows a second lobe about 10 dB below the main lobe, corresponding to a reflection from a building close to the roof top receiver.

Distributions of measured azimuth gains in both roof-edge and offset measurements are plotted in Fig. 11. Colored regions around the roof-edge and offset cases are 90% confidence intervals [27]. In the roof-edge case 90% of observed azimuth gains are within 2 dB of the antenna nominal azimuth gain, a reference measured in an anechoic chamber. When the base antenna is offset by 5 m from the roof edge, 90% of observed azimuth gains are within 4 dB of the nominal gain. The additional degradation may be due to scattering from nearby roof structures.

Degradation from the nominal 14.5 dB azimuth gain observed in Fig. 11 may be used to account for gain degradation in link budget calculations, as well as to derive corresponding azimuth angle spread. We note that the gain degradation found in these channels is small, implying narrow angle spread and supporting effectiveness of using high gain base antennas in street canyons. In contrast, high directional antennas are ineffective in fully scattering channels where power versus angle is constant on average, although the angular spectrum instantiation is subject to direction-dependent fading. As a result, modest diversity gains are achievable by selecting the direction with the highest power instantiation, as shown in Fig. 11 where the simulated arrivals from different directions follow the i.i.d. complex Gaussian distribution, as appropriate in full scattering. The amplitude of the complex sum is then Rayleigh distributed. The complex channel spectrum is convolved with the complex antenna pattern [23] to generate instantiations of the pattern, whose gain is computed using (6) and plotted as the "full scattering" distribution in Fig. 11.

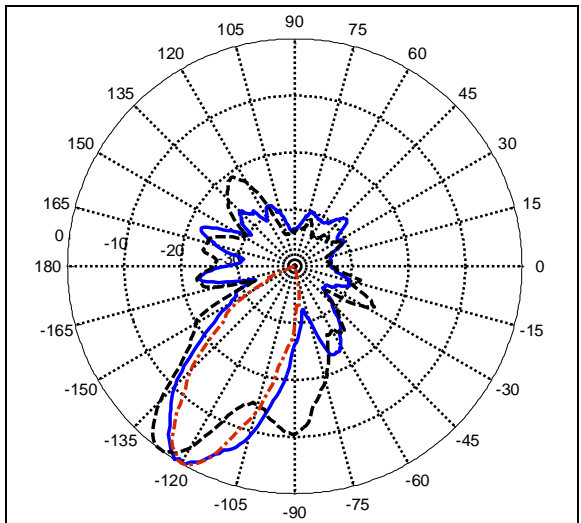

Figure 10. Measured normalized patterns for base in direct view of the street. Solid blue is measured at 580 m, dashed black at 100 m. Red dash-dot pattern was measured in anechoic chamber.

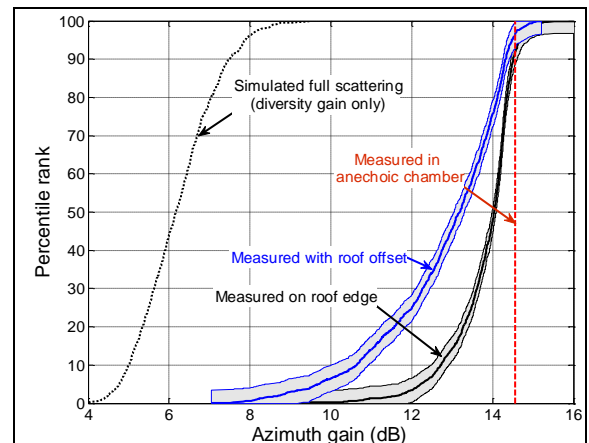

Figure 11. Distribution of observed effective azimuth gain for base antennas on the roof edge and offset from roof edge.

---

[2] The $10°$ (24 dBi) horn at the base has a beam pattern that is close to some of the early 28 GHz phased array products used in base stations.

## V. ACHIEVABLE RATES BASED ON PROPAGATION MEASUREMENTS

To evaluate coverage in dense urban deployment, we simulate rates in an idealized urban network with 200 m by 50 m rectangular city blocks and with cells placed at street corners separated by 400 m along a street, as illustrated in Fig. 12, where blue stars indicate base station sites. To provide coverage along all streets, each site is located at the intersection, with four cells covering four directions along the streets, but not all streets contain a base station. We use path loss formulas presented above for same-street (roof-edge) (1) and around-the-corner channels (3).

We focus on the downlink (DL) cell rate assuming 800 MHz bandwidth at 28 GHz band. Each cell, mounted on the roof of a 20 m high corner building is assumed to have 28 dBm transmit power and 23 dBi nominal antenna gain. Each UE is assumed to have 6 dBi antenna gain with noise figure of 9 dB. Path loss models are from Section III and gain degradation is computed based on our measurements in Sec. IV. The base station is assumed to aim towards the UE it is serving and interference from neighboring cells is included in SINR calculation. Since UEs are served by the strongest base, they may benefit from macro diversity. Rates are computed as the Shannon rate of the $10^{th}$ percentile DL SINR with a 3 dB implementation penalty. The resulting coverage map is illustrated in Fig. 12.

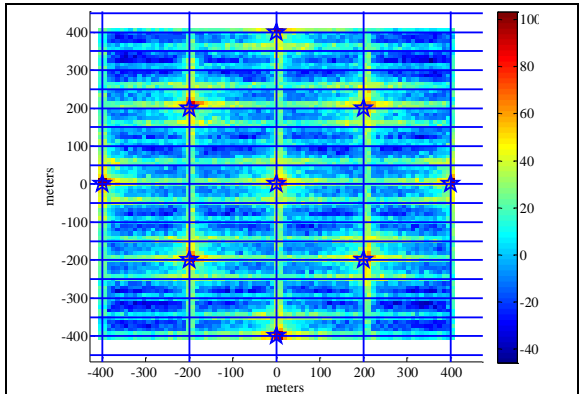

Figure 12. Simulated SNR distribution on urban streets with bases (blue stars) placed at intersections, separated 400 m along a street, 12 sites/sq km.

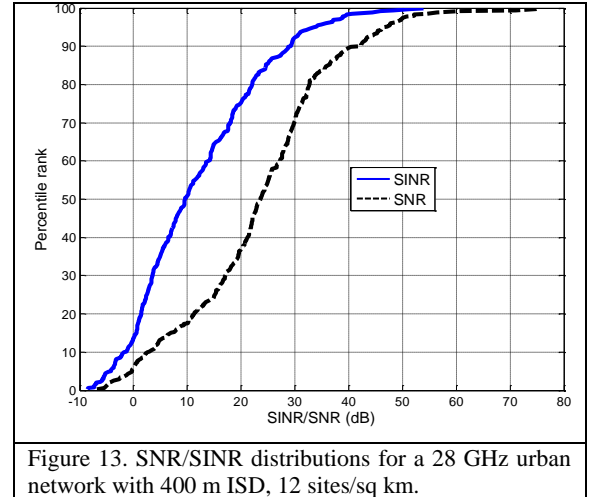

Figure 13. SNR/SINR distributions for a 28 GHz urban network with 400 m ISD, 12 sites/sq km.

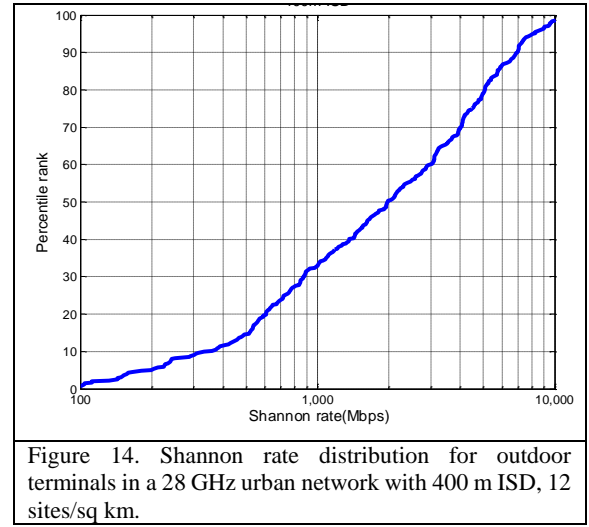

Figure 14. Shannon rate distribution for outdoor terminals in a 28 GHz urban network with 400 m ISD, 12 sites/sq km.

The SNR/SINR and rate distributions are shown in Fig. 13 and Fig. 14, respectively. It may be observed in Fig. 13 that the SNR is 5 dB higher than SINR for 90% of outdoor locations, indicating an interference-limited (outdoor) system. The $10^{th}$ percentile rate for 400m inter-site-distance (ISD) is 350 Mbps for outdoor terminals. Higher cell densities of 25 and 50 sites/sq km, not shown here due to space limitation, were found to have similar rate distributions as 12 sites/sq km shown in Fig. 14. Naturally, actual user rates are impacted by the cell density as it determines the degree of sharing of bandwidth among users. Should a different directional gain degradation model or around-the-corner path loss model be used, or the base be offset from roof edge, the results might be different. Their impact on system performance is an interesting direction for future work.

## VI. Discussion and conclusions

Over 21 million CW power samples in 3000 links on 12 streets in Manhattan, NY and Valparaíso, Chile were collected to characterize propagation at 28 GHz in urban canyons from roof top base stations using a vertically polarized azimuthally rotating horn antenna at the base, at ranges up to 800 m.

Slope-intercept fit to same-street path gain data from roof-edge base is found to have RMS deviation of 7.1 dB. Large statistical significance of the data allowed for high 90% certainty in fit parameters (+/- 2.7 dB for 1-m intercept and +/- 0.12 for slope).

Standard 3GPP models were found to produce 12 to 17 dB RMS loss relative to our data. Measured path gain was found to suffer excess loss relative to free space that increased with distance, reaching 11 dB at 200 m. It was found that standard ray tracing in this simple same-street scenario overpredicts signal strength (13 dB at 200 m), probably due to the omission of difficult-to-model scatter from street objects, such as vehicles, pedestrians and trees. Separate fits to subsets of data collected in Manhattan and Valparaíso were within 2 dB of each other, as were lamppost to street and roof to street. Offsetting the base antenna 5 m away from the roof edge, towards the center of the building led to an additional average loss of 15 dB at 100 m. Around the corner propagation is well modeled by a diffraction-based model using an empirical diffraction coefficient of 2 dB.

90% of measured effective azimuthal base antenna gains were within 2 dB of nominal, indicating low angle spread compared to the nominal antenna beamwidth of 10°.

Simulation of network performance for outdoor users indicated 90% users have a Shannon rate of 350 Mbps or higher, with 400 m ISD and 12 sites/sq km. For the simulated site arrangement, the majority of outdoor locations were not in line of sight to any cell. Quadrupling site density led to near quadrupling of the shared user rate, consistent with decreased number of users per cell.

## Acknowledgment

The authors wish to acknowledge the support received from the Chilean Research Agency ANID, through research grants ANID PIA/APOYO AFB180002, ANID FONDECYT/INICACION 11171159, ANID REDES 180144, Project VRIEA-PUCV 039.430/2020 and Project VRIEA-PUCV 039.437/2020. The work of T. Chen, M. Kohli and G. Zussman was supported in part by NSF grants ECCS-1547406 and CNS-1827923, and NSF-BSF grant CNS-1910757. Many thanks to Hector Carrasco, Leonardo Guerrero and Rene Pozo for designing and building the platform, Cuong Tran for essential diagnostics and repair, Tianyi Dai, Angel Daniel Estigarribia and Shounak Roy for assistance in measurements.
## References

[1] M. Kyro, V. Kolmonen and P. Vainikainen, "Experimental Propagation Channel Characterization of mm-Wave Radio Links in Urban Scenarios," IEEE Antennas and Wireless Propagation Letters, vol. 11, pp. 865-868, Aug. 2012.

[2] T. S. Rappaport, F. Gutierrez, E. Ben-Dor, J. N. Murdock, Y. Qiao and J. I. Tamir, "Broadband Millimeter-Wave Propagation Measurements and Models Using Adaptive-Beam Antennas for Outdoor Urban Cellular Communications," IEEE Transactions on Antennas and Propagation, vol. 61, pp. 1850-1859, Apr. 2013.

[3] H. J. Thomas, R. S. Cole and G. L. Siqueira, "An experimental study of the propagation of 55 GHz millimeter waves in an urban mobile radio environment," IEEE Transactions on Vehicular Technology, vol. 43, pp. 140-146, Feb. 1994.

[4] V. Semkin, U. Virk, A. Karttunen, K. Haneda and A. V. Räisänen, "E-band propagation channel measurements in an urban street canyon," 2015 9th European Conference on Antennas and Propagation (EuCAP), Lisbon, Portugal, Apr. 2015.

[5] H. Masui, T. Kobayashi and M. Akaike, "Microwave path-loss modeling in urban line-of-sight environments," IEEE Journal on Selected Areas in Communications, vol. 20, pp. 1151-1155, Aug. 2002.

[6] R. J. Weiler, M. Peter, T. Kühne, M. Wisotzki and W. Keusgen, "Simultaneous millimeter-wave multi-band channel sounding in an urban access scenario," 2015 9th European Conference on Antennas and Propagation (EuCAP), Lisbon, Portugal, Apr. 2015.

[7] A. F. Molisch et al., "Millimeter-wave channels in urban environments," 2016 10th European Conference on Antennas and Propagation (EuCAP), Davos, Switzerland, Apr. 2016.

[8] W. Keusgen, R. J. Weiler, M. Peter, M. Wisotzki and B. Göktepe, "Propagation measurements and simulations for millimeter-wave mobile access in a busy urban environment," 2014 39th International Conference on Infrared, Millimeter, and Terahertz waves (IRMMW-THz), Tucson, AZ, Sept. 2014.

[9] A. Karttunen, A. F. Molisch, S. Hur, J. Park, C. J. Zhang, "Spatially Consistent Street-by-Street Path Loss Model for 28-GHz Channels in Micro Cell Urban Environments", IEEE Transactions on Wireless Communications, vol.16, pp. 7538-7550, Nov. 2017.

[10] K. Haneda, N. Omaki, T. Imai, L. Raschkowski, M. Peter and A. Roivainen, "Frequency-agile pathloss models for urban street canyons," IEEE Transactions on Antennas and Propagation, vol. 64, pp. 1941-1951, May 2016.

[11] K. Haneda et al., "5G 3GPP-like channel models for outdoor urban microcellular and macrocellular environments," IEEE Vehicular Technology Conference (VTC Spring), Nanjing, China, May 2016.

[12] T. S. Rappaport, Y. Xing, G. R. MacCartney, A. F. Molisch, E. Mellios and J. Zhang, "Overview of millimeter wave communications for fifth-generation (5G) wireless networks-with a focus on propagation models," IEEE Transactions on Antennas and Propagation, vol. 65, pp. 6213-6230, Dec. 2017.

[13] J. Ko, Y.-J. Cho, S. Hur, T. Kim, J. Park, A. F. Molisch, K. Haneda, M. Peter, D. Park, D.-H. Cho, "Millimeter-Wave Channel Measurements and Analysis for Statistical Spatial Channel Model in In-Building and Urban Environments at 28 GHz", IEEE Transactions on Wireless Communications, vol. 16, pp. 5853-5868, Sept. 2017.

[14] V. Raghavan, A. Partyka, L. Akhoondzadeh-Asl, M. A. Tassoudji, O. H. Koymen and J. Sanelli "Millimeter wave channel measurement and implications for PHY layer design", IEEE Transactions on Antennas and Propagation, vol. 65, pp. 6521-6533, Dec. 2017.

[15] H. Zhao, R. Mayzus, S. Sun, M. Samimi, J. K. Schulz; Y. Azar, K. Wang, G. N. Wong, F. Gutierrez, T. S. Rappaport, "28 GHz

millimeter wave cellular communication measurements for reflection and penetration loss in and around buildings in New York city," IEEE International Conference on Communications (ICC), Budapest, Hungary, Jun. 2013.
[16] R. Wang, C. U. Bas, S. Sangodoyin, S. Hur, J. Park, J. Zhang and A. F. Molisch, "Stationarity region of Mm-Wave channel based on outdoor microcellular measurements at 28 GHz," IEEE Military Communications Conference (MILCOM), Baltimore, MD, Oct. 2017.
[17] J. Lee, M. Kim, J. Liang, J. Park and B. Park, "Frequency range extension of the ITU-R NLOS path loss models applicable for urban street environments with 28 GHz measurements," 2016 10th European Conference on Antennas and Propagation (EuCAP), Davos, Switzerland, Apr. 2016.
[18] C. Larsson, B. Olsson and J. Medbo, "Angular Resolved Pathloss Measurements in Urban Macrocell Scenarios at 28 GHz," 2016 IEEE 84th Vehicular Technology Conference (VTC-Fall), Montreal, Canada, Sept. 2016.
[19] S. Hur et al., "Proposal on Millimeter-Wave Channel Modeling for 5G Cellular System," IEEE Journal of Selected Topics in Signal Processing, vol. 10, pp. 454-469, Apr. 2016.
[20] "Study on channel model for frequencies from 0.5 to 100 GHz," 3GPP Technical Report TR 38.901 v14.1.1, Jul. 2017. http://www.3gpp.org/ftp/specs/archive/38_series/38.901/38901-e11.zip
[21] "Further advancements for E-UTRA physical layer aspects (Release 9)", 3GPP Technical Report TR 36.814, V9.2.0, Mar. 2017.
[22] "Guidelines for evaluation of radio interface technologies for IMT-Advanced," ITU-R Report M.2135-1, Dec. 2009.
[23] D. Chizhik, J. Du, R. Feick, G. Castro, M. Rodriguez and R. A. Valenzuela, "Path Loss and Directional Gain Measurements at 28 GHz for non-line-of-sight coverage of indoors with corridors," IEEE Transactions on Antennas and Propagation, vol. 68, pp. 4820-4830, Jun. 2020.
[24] J. Du, D. Chizhik, R. Feick, M. Rodríguez, G. Castro and R. A. Valenzuela, "Suburban Fixed Wireless Access Channel Measurements and Models at 28 GHz for 90% Outdoor Coverage," IEEE Transactions on Antennas and Propagation, vol. 68, pp. 411-420, Jan. 2020.
[25] "Propagation data and prediction methods for the planning of short-range outdoor radio communication systems and radio local area networks in the frequency range 300 MHz to 100 GHz," Recommendation ITU-R P.1441-9, Jun. 2017.
[26] M. Sasaki, W. Yamada, T. Sugiyama, M. Mizoguchi and T. Imai, "Path loss characteristics at 800 MHz to 37 GHz in urban street microcell environment," 2015 9th European Conference on Antennas and Propagation (EuCAP), Lisbon, Portugal, Apr. 2015.
[27] A. Dvoretzky; J. Kiefer, J. Wolfowitz, (1956). "Asymptotic minimax character of the sample distribution function and of the classical multinomial estimator," Annals of Mathematical Statistics, vol. 27, pp. 642-669, Sept. 1956.
[28] H. L. Bertoni, Radio Propagation for Modern Wireless Systems, Prentice Hall, 2000.
[29] J. S. Lu, H. L. Bertoni, K. A. Remley, W. F. Young and J. Ladbury, "Site-Specific Models of the Received Power for Radio Communication in Urban Street Canyons," IEEE Transactions on Antennas and Propagation, vol. 62, pp. 2192-2200, Apr. 2014.
[30] D. Dupleich, R. Müller, S. Skoblikov, J. Luo, G. Del Galdo and R. S. Thomä, "Multi-band Double-directional 5G Street Canyon Measurements in Germany," 2019 European Conference on Networks and Communications (EuCNC), Valencia, Spain, Jun. 2019.
[31] D. Dupleich et al., "Multi-band Propagation and Radio Channel Characterization in Street Canyon Scenarios for 5G and Beyond," IEEE Access, vol. 7, pp. 160385-160396, Oct. 2019.